# Evidence for strong localization of orbital polarization

Taiyang Zhang,[1,2] Lujun Zhu,[3] Zhihao Yan,[1,2] Lijun Zhu[1,2]*
[1] *State Key Laboratory of Semiconductor Physics and Chip Technologies, Institute of Semiconductors, Chinese Academy of Sciences, Beijing 100083, China*
[2] *Center of Materials Science and Optoelectronics Engineering, University of Chinese Academy of Sciences, Beijing 100049, China*
[3] *College of Physics and Information Technology, Shaanxi Normal University, Xi'an 710062, China*
**ljzhu@semi.ac.cn*

Whether orbital polarization propagates has become the most essential question of the blooming orbitronics that aims to generate non-local orbital current, torque, and pumping. Recent theories have suggested a strong orbital Hall effect within the light metal Al and a strong orbital Rashba effect at Co/Al interfaces, providing ideal platforms for experimental verification of possible orbital transport effects. Here, we report robust experimental evidence for the strong localization of orbital polarization. We demonstrate that neither the bulk nor the interface of the Al contributes a detectable orbital torque on adjacent magnetic layer with strong bulk and interfacial spin-orbit coupling necessary for potential orbital-spin conversion. These results reveal that orbital polarization undergoes much faster relaxation than spin polarization and hardly participates in non-local accumulation, transport, or torque generation due to quenching and spin-orbit coupling. The strong localization of orbital polarization is a unique groundbreaking conceptional advance towards solving the essential debate on orbitronics.

**Introduction.** Employment of non-local spin polarization into electronics inaugurated spintronics. In the past two decades, spin polarization was consensually known to be converted from *local* orbital polarization via spin-orbit coupling (SOC) in the case of the spin Hall effect (SHE)[1], while orbital polarization can hardly escape from its host material for interlayer transport. While orbital polarization cannot interact with magnetization [2], non-local spin polarization can generate strong spin-orbit torques (SOTs) on nanomagnets (Fig. 1a) for ultrafast energy-efficient magnetic memory and computing [3–14]. Recently, there is a blooming interest in revisiting the fundamental question as to whether orbital polarization *propagated* and created torque and pumping effects in adjacent materials [15–40].

The key assumption of the orbitronics concepts of orbital torque and pumping is the long-range transport of orbital polarization. However, it has remained unsettled as to whether and how orbital polarization, if generated by the bulk orbital Hall effect or interfacial orbital Rashba effect, could indeed accumulate at the surface of its host material or be injected into and from an adjacent layer via non-local transport. Some theories [1,41,42] have predicted that the orbital polarization has extremely short, atomic scale relaxation length in the host material (0.2-0.4 nm) mainly due to ultrafast quenching by crystal field, regardless of the SOC. Orbital-spin conversion, the second important orbital relaxation mechanism, is negligible unless the host material had a high spin Hall ratio such that the orbital-spin conversion manifests as the SHE [1,43]. Experimentally, the heavy metal Ta with giant theoretically predicted orbital Hall conductivity is clarified to generate no torque other than the spin Hall torque on adjacent ferromagnets [44]. Physical origins alternative to orbital polarization have been identified for some experimental data features claimed to imply occurrence of orbital transport in magnetic heterostructures (e.g., unusual magnetoresistance [45], the magnet-type-dependent spin torque ferromagnetic resonance signal [44] and terahertz emission [46]).

Given the fundamental importance for the entire field of spintronics, high-precision experiments are urgently required to clarify whether orbital polarization propagates. Recently, the light metal Al with negligible SOC and spin Hall ratio has attracted considerable interest as a potential orbital current emitter [26] for the theoretical prediction of high orbital Hall conductivity of -700 ($\hbar$/e) $\Omega^{-1}$ cm$^{-1}$ [16] and strong orbital Rashba effect at the Co/Al interface [28,29], providing ideal platforms for experimental clarification of whether orbital polarization is localized.

Here, we for the first time report robust experimental evidence that no orbital torque can be generated due to the orbital Hall effect or the orbital Rashba effect in Pt/Co/Al and FePt/Al heterostructures, despite the strong SOC at the Co/Al interface and in the FePt. These results reveal that even the light metals with negligible SOC and spin Hall ratio forbid long-range accumulation and transport of orbital polarization due to the ultrafast orbital quenching.

**Sample details.** For this work, Pt 5/Co 2/Al 0-4 trilayers and FePt 5/Al 0-4 bilayers (numbers are the layer thicknesses in nm) are fabricated to test potential orbital transport and orbital torque due to the orbital Hall effect and orbital Rashba effect of Al and its interface. Each sample is sputter-deposited on thermally oxidized Si substrates at room temperature and protected by a MgO 1.6/Ta 1.5 bilayer that is fully oxidized upon exposure to the atmosphere [47]. A 1 nm amorphous Ta was deposited as the adhesion layer for the Pt/Co/Al samples but not for the FePt/Al samples. The cross-sectional transmission electron microscopy imaging results reveal reasonably smooth interfaces for the Pt/Co/Al and FePt/Al samples (Fig. 1b). These samples are patterned into 5 μm × 60 μm Hall bars by photolithography and argon ion milling, followed by the sputter-deposition of Ti 5/Pt 150 contacts (Fig. 1c). The in-plane magnetization hysteresis (Fig. 1d) and the planar Hall voltage (Fig. 1e) measurements reveal in-plane magnetic anisotropy and saturation magnetization ($M_s$) of 1255 emu/cm$^3$ for the Co and 670 emu/cm$^3$ for the FePt, which are consistent with our previous reports [44,48].

**Spin-orbit torque characterizations.** The dampinglike and fieldlike SOTs of the Pt/Co/Al and the FePt/Al are quantified from angle-dependent harmonic Hall voltage measurements with the thermal effects carefully taken into account [11,49–53]. While a sinusoidal electric field ($E$) is applied onto the Hall bar (Fig. 1c), the first and second harmonic Hall voltages ($V_{1\omega}$ and $V_{2\omega}$) are collected as a function of the angle ($\varphi$) of the in-plane magnetic field ($H_{xy}$) relative to $E$. The magnitude of $H_{xy}$ is varied between 0.75 kOe and 3.75 kOe to assure the macrospin behavior (Fig. 1d)[54] and negligible ordinary Nernst effect [55]. When an IMA macrospin sample interacts with a transversely polarized spin current, $V_{2\omega}$ reads [51]

$$V_{2\omega} = V_{\mathrm{DL+ANE}}\cos\varphi + V_{\mathrm{FL+Oe}}\cos\varphi\cos2\varphi + V_{\mathrm{PNE}}\sin2\varphi, \quad (1)$$

with

$$V_{\mathrm{DL+ANE}} = V_{\mathrm{AHE}}H_{\mathrm{DL}}/2(H_{xy}-H_{\mathrm{k}}) + V_{\mathrm{ANE}}, \quad (2)$$

$$V_{\mathrm{FL+Oe}} = -V_{\mathrm{PHE}}(H_{\mathrm{FL}}+H_{\mathrm{Oe}})/H_{xy}. \quad (3)$$

Here, $H_{\mathrm{DL}}$ is the dampinglike SOT field, $H_{\mathrm{FL}}$ is the fieldlike SOT field, $H_{\mathrm{Oe}}$ is the transverse Oersted field exerted on the magnetic layer by the in-plane charge current in other layers, $H_{\mathrm{k}}$ is the effective PMA field (negative for IMA samples), $V_{\mathrm{AHE}}$ is the anomalous Hall voltage, $V_{\mathrm{ANE}}$ is the anomalous Nernst voltage induced by the vertical thermal gradient, $V_{\mathrm{PNE}}$ is the planar Nernst voltage induced by the longitudinal thermal gradient [52], and $V_{\mathrm{PHE}}$ is the planar Hall voltage ($V_{1\omega} = V_{\mathrm{PHE}}\sin2\varphi$ under zero $H_z$, Fig. 1e).

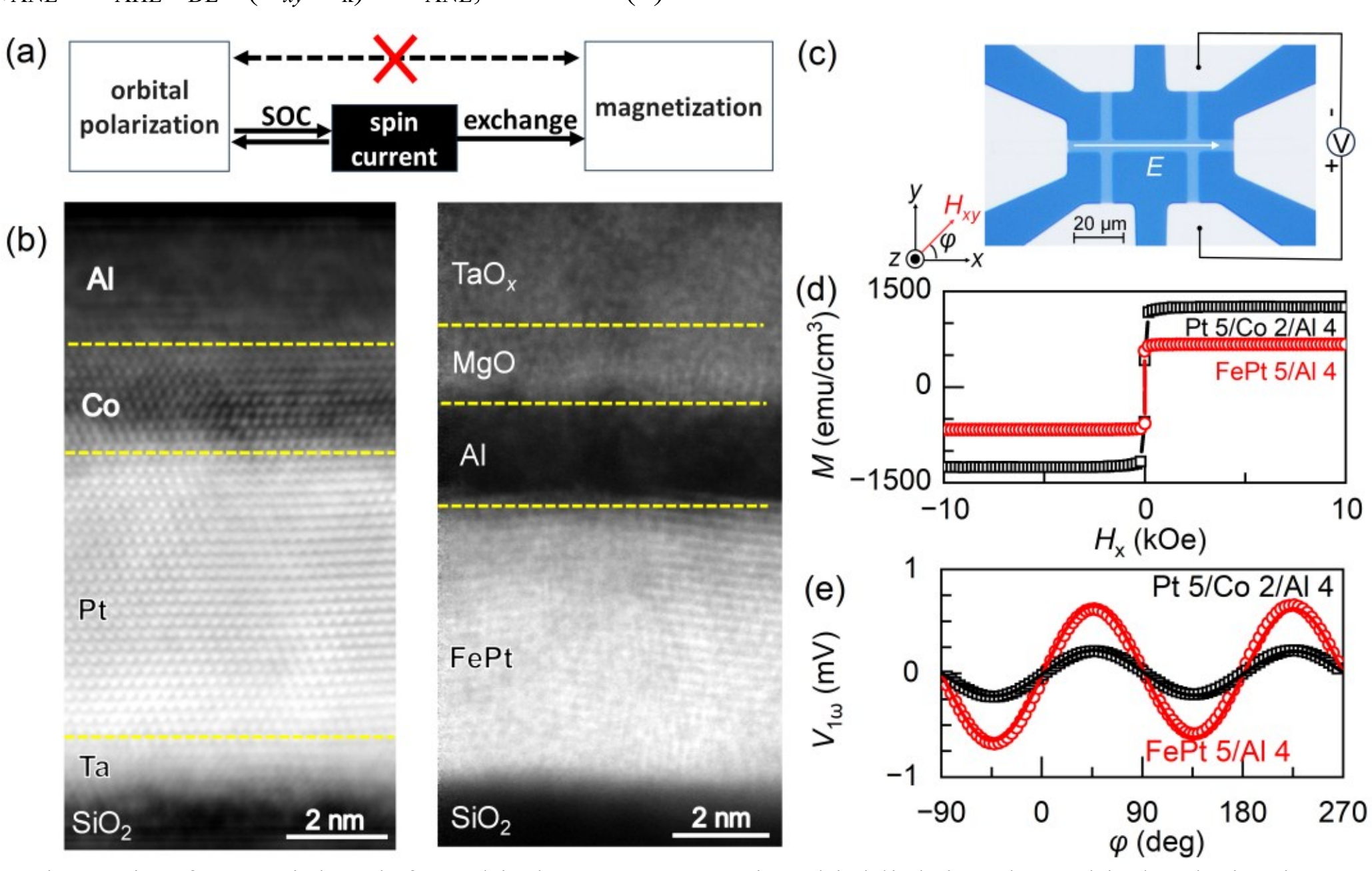

**Figure 1** (a) Schematic of potential path for orbital torque generation, highlighting that orbital polarization cannot interact with magnetization unless being converted into spin current via SOC. (b) Cross-sectional transmission electron microscopy images of a Pt/Co/Al sample and FePt/Al samples. (c) Optical microscopy image of the Hall bar device and the measurement coordinate. (d) Magnetization hysteresis and (e) First harmonic Hall voltage vs the azimuth angle ($\varphi$) of the in-plane magnetic field ($H_{xy}$ = 3.5 kOe) for the Pt 5/Co 2/Al 4 and the FePt 5/Al 4.

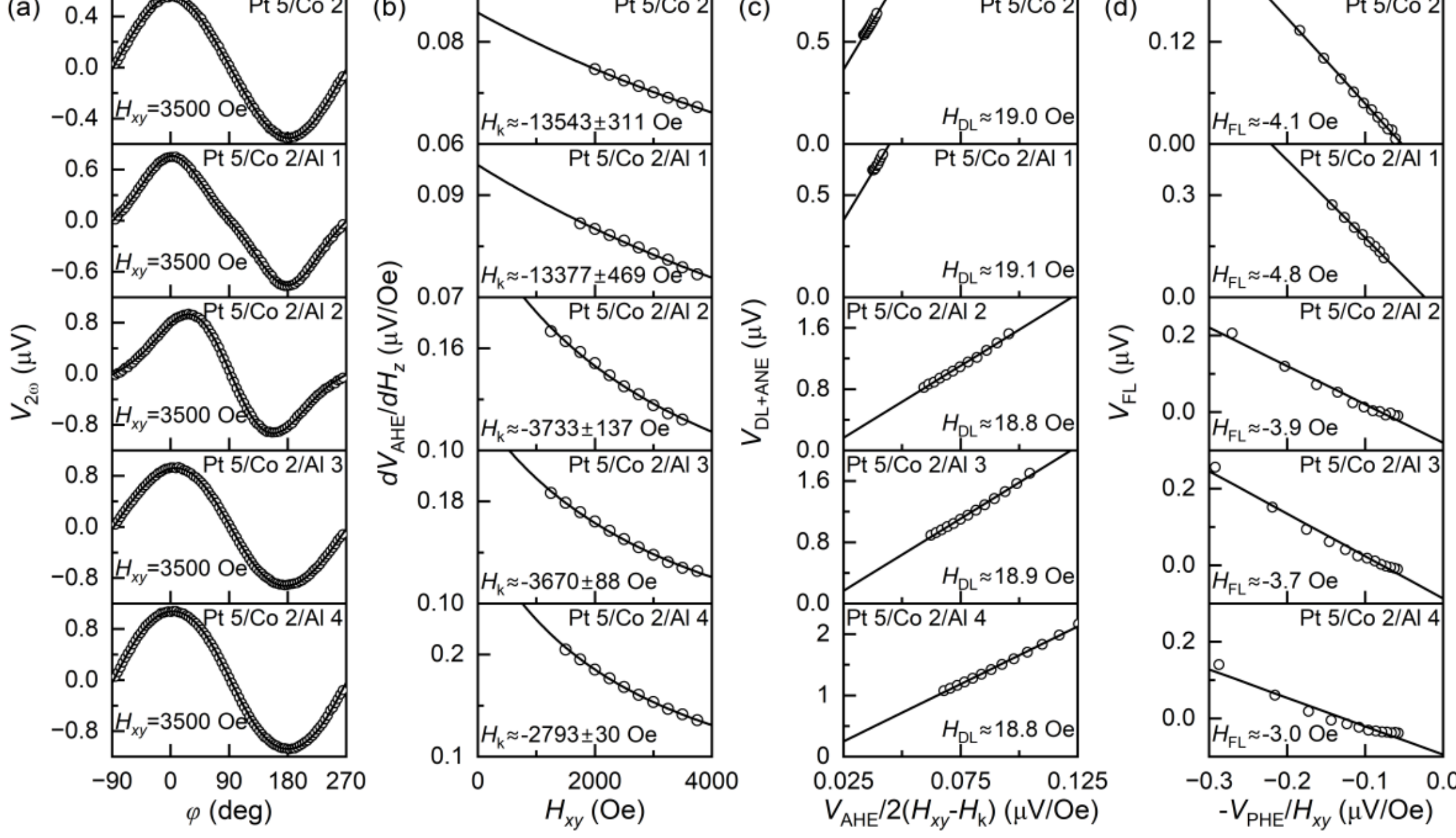

**Figure 2. Harmonic Hall voltage analysis of the Pt/Co/Al devices**. (a) $V_{2\omega}$ vs $\varphi$ ($H_{xy}$ = 3.5 kOe), (b) $dV_{\mathrm{AHE}}/dH_z$ vs $H_{xy}$, (c) $V_{\mathrm{DL+ANE}}$ vs $V_{\mathrm{AHE}}/2(H_{xy}-H_{\mathrm{k}})$, and (d) $V_{\mathrm{FL+Oe}}$ vs $-V_{\mathrm{PHE}}/H_{xy}$ for the FePt/Al bilayers with Al thickness of 0 nm, 1 nm, 2 nm, 3nm, and 4 nm, respectively. The solid lines in (a)-(d) represent the best fits of the data to Eq. (1), Eq. (4), Eq. (2), and Eq. (3), respectively. The electric field $E$ is 33.3 kV/m.

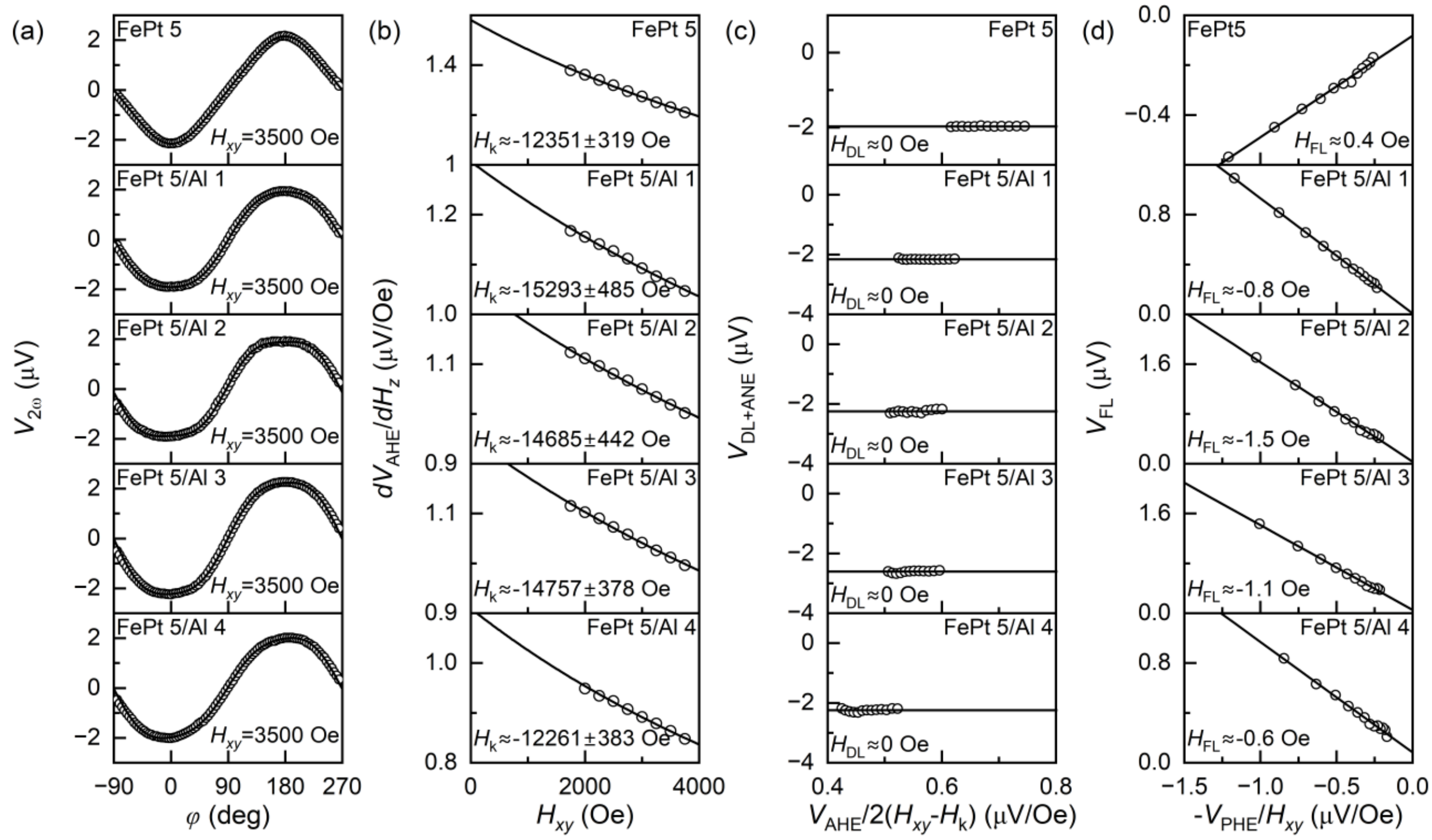

**Figure 3. Harmonic Hall voltage analysis of the FePt/Al bilayers**. (a) $V_{2\omega}$ vs $\varphi$ ($H_{xy}$ = 3.5 kOe), (b) $dV_{AHE}/dH_z$ vs $H_{xy}$, (c) $V_{DL+ANE}$ vs $V_{AHE}/2(H_{xy}$-$H_k)$, and (d) $V_{FL+Oe}$ vs -$V_{PHE}/H_{xy}$ for the FePt/Al bilayers with Al thickness of 0 nm, 1 nm, 2 nm, 3nm, and 4 nm, respectively. The solid lines in (a)-(d) represent the best fits of the data to Eq. (1), Eq. (4), Eq. (2), and Eq. (3), respectively. The electric field $E$ is 66.7 kV/m.

For clarity, we show the harmonic Hall voltage measurement data of the Pt/Co/Al devices in Fig. 2a-d and the FePt/Al devices in Fig. 3a-d. The values of $V_{DL+ANE}$ and $V_{FL+Oe}$ for each magnitude of $H_{xy}$ are determined from the fits of $V_{2\omega}$ vs $\varphi$ to Eq. (1) (Fig. 2a and Fig.3a). As shown in Fig. 2b and Fig.3b, the values of $H_k$ and $V_{AHE}$ are determined from the fits of $dV_{1\omega}/dH_z$ vs $H_{xy}$ to the relation [13,56,57]

$$dV_{1\omega}/dH_z = V_{AHE}/(H_{xy}-H_k), \tag{4}$$

where $dV_{1\omega}/dH_z$ is measured from the linear scaling of $V_{AHE}$ with the swept out-of-plane field ($H_z$) under given in-plane magnetic field $H_x$ . As shown in Fig. 2c,d and Fig. 3c,d, the values of $H_{DL}$ and $H_{FL}$ are estimated from the slopes of the linear fits of $V_{DL+ANE}$ vs $V_{AHE}/2(H_{xy}$-$H_k)$ to Eq. (2) and of $V_{FL+Oe}$ vs -$V_{PHE}/H_{xy}$ to Eq. (3), respectively. $H_{Oe}$ is estimated as $(j_{Pt}d_{Pt}$-$j_{Al}d_{Al})/2$ for the Pt/Co/Al devices and as -$j_{Al}d_{Al}/2$ for the FePt/Al devices. Here, $j_{Pt(Al)}$ and $d_{Pt}(d_{Al})$ are the current density and the thickness of the Pt (Al) layers.

With the values of $H_{DL}$ and $H_{FL}$ of the Pt/Co/Al and the FePt/Al, the dampinglike and fieldlike SOT efficiencies per electric field ($\xi_{DL}^{E}$ and $\xi_{FL}^{E}$) are estimated following

$$\xi_{DL(FL)}^{E} = (2e/\hbar)\mu_0 M_s t_{FM} H_{DL(FL)}/E, \tag{5}$$

where $e$ is the elementary charge, $\hbar$ the reduced Planck's constant, $\mu_0$ the permeability of vacuum, and $t_{FM}$ the total thickness of the magnetic layer.

**Absence of orbital torque.** As plotted in Fig. 4a, $\xi_{DL}^{E}$ for the Pt/Co/Al remains essentially at the same high value of (4.33±0.03)×$10^5$ $\Omega^{-1}$ $m^{-1}$ as the Al thickness ($d_{Al}$) is varied from 0 nm to 4 nm, which reveals negligible torque contribution from the theoretically predicted bulk ordinary Hall effect in the Al layer and the orbital Rashba effect of the Co/Al interface. The fieldlike torque of $\xi_{FL}^{E}$ = (0.89±0.15)×$10^5$ $\Omega^{-1}$ $m^{-1}$ is much weaker than and of the opposite sign compared to the dampinglike torque. This observation is consistent with the SHE of the Pt being the only origin of the observed dampinglike and fieldlike torques [10,11,14]. The slight variation of the fieldlike torque with the Al thickness and thus the details of the Co/Al interfaces, which is consistent with previous report of sensitivity of the fieldlike spin Hall torque to the interface details [58]. The slight variation of the fieldlike torque with the Al thickness seems unlikely to be attributed to any spin or orbital Rashba effect at the Co/Al interface because the interfacial torque should not vary much at large Al thicknesses of 4 and 5 nm. For the FePt/Al, both the dampinglike and fieldlike torque are negligibly small for all the Al thicknesses (0-4 nm). The FePt results reveal that the orbital Hall effect of the Al generates no torques on the adjacent magnetic layer within the experimental resolution even if the bulk SOC of the magnetic layer is very strong. The absence of orbital torque in the Al/ferromagnet heterostructures is in good agreement with previous reports on Ta/ferromagnet bilayers [44]. The observation of negligible self-induced SOT within the 5 nm FePt is consistent with our previous reports [44,48].

**Strong localization of orbital polarization.** Microscopically, the absence of orbital torque in the Pt/Co/Al and FePt/Al indicate negligible injection of orbital current (flow of orbital polarization) into the magnetic layers (Co and FePt) from the Al and its interface (see the insets of Fig. 4a). This is because an orbital current, if significant, would be partially converted into spin current by the SOC of the interface and the magnetic layer and be detected in the SOT experiments. Note that Co has a significant bulk SOC (84 meV)[59] that were claimed as efficient orbital detectors [26,27], while FePt has strong bulk SOC of 335 meV [44], which is more than a factor of 3 stronger than that of Co. The Co/Al also has strong interfacial SOC at the Co/Al interface. As indicated by the Bruno's model [60] and previous experiments [57,61], the SOC of a magnetic interface can be parameterized using the PMA energy density ($K_s$) of the interface. Following the relation of $H_k = 4\pi M_s - 2K_s/M_s t_{Co}$, $K_s$ of the two Co interfaces is estimated to be 0.28±0.04 erg/cm$^2$ for the Pt/Co/MgO, be 0.30±0.06 erg/cm$^2$ for Pt/Co/Al 1 to 1.55±0.07 for Pt/Co/Al 2-4. Assuming an approximately constant PMA for the bottom Pt/Co interface of the Pt/Co/Al samples, $K_s$ for the Co/Al interface would be greater than ≈1.2 erg/cm$^2$, which is giant compared to that of typical magnetic interfaces [13,57].

The absence of orbital current injection into the magnetic layers from the Al and its interface most likely suggests ultrafast orbital quenching in the Al. We propose in Fig. 4b that orbital polarization, once generated, would relax rapidly within the host orbital Hall metal via ultrafast quenching due to the crystal field and via orbital-spin conversion due to SOC (in the case of a high spin Hall ratio)[1]. Consequently, the ultrafast relaxation of orbital polarization prevents any non-local orbital transport including the accumulation and interfacial diffusion. For the similar reason, we propose in Fig.4c that, even when a strong orbital current is somehow injected into a thin-film magnet, it should relax mainly via ultrafast quenching without relevance to torque generation, leaving only a small portion being converted into spin polarization only if the magnet had a high spin-Hall ratio.

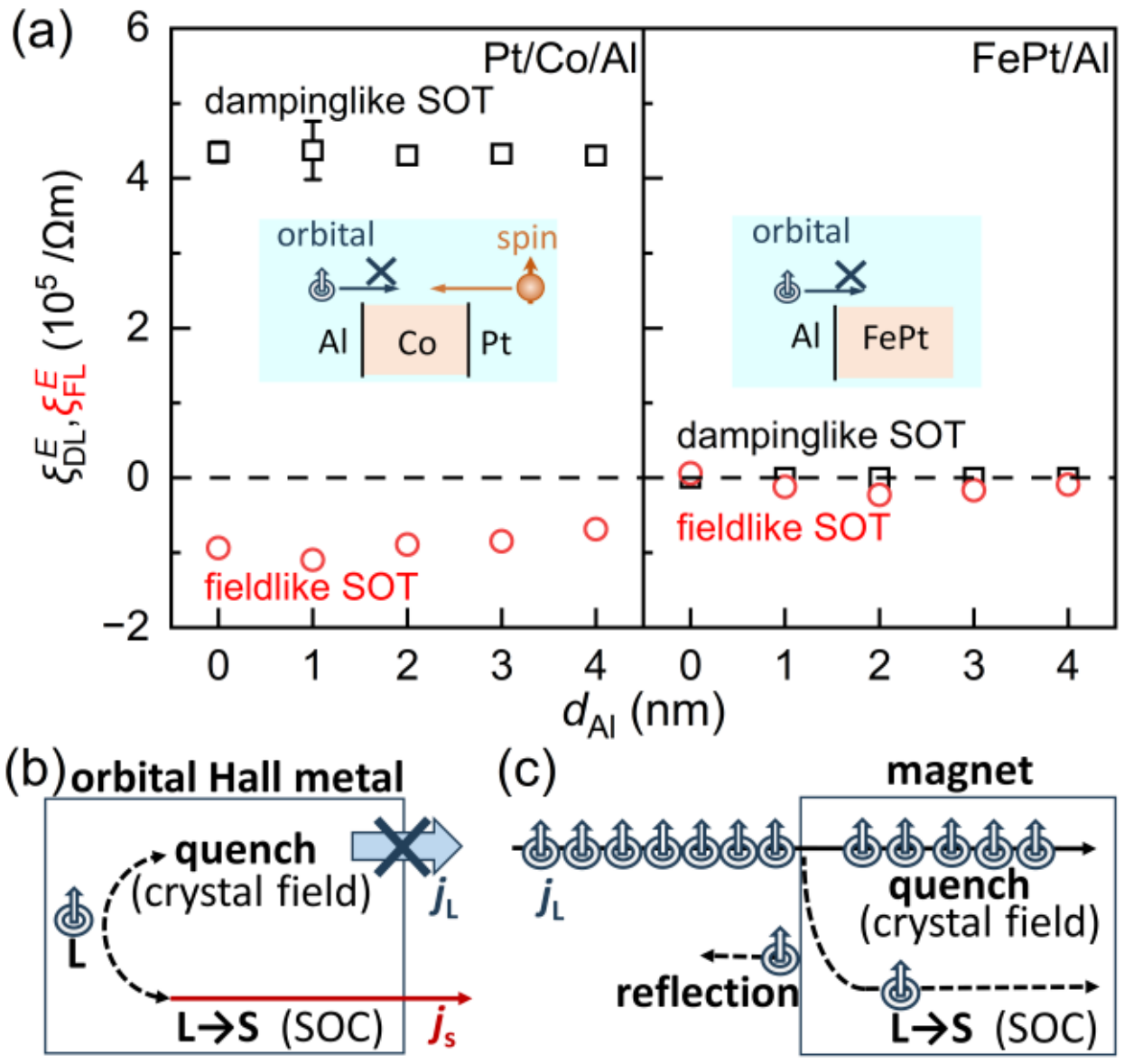

**Figure 4** (a) Dampinglike and fieldlike SOT efficiencies per electric field. The dashed line is to guide the eyes. Error bars are standard deviations. Insets suggest the orbital and spin transport into the magnetic layer. (b) Relaxation of orbital polarization within the orbital Hall metal. (c) Relaxation of an incident orbital current within a magnet.

**Conclusion.** We have demonstrated robust experimental evidence that no orbital current torque is generated due to the Al or Co/Al interface despite the theoretically predicted bulk orbital Hall effect and interfacial orbital Rashba effect. These findings suggest that orbital polarization undergoes ultrafast relaxation (likely via orbital quenching) that prevents effective non-local transport from accumulation within the light-metal generator (Al and Co/Al interface in this work) and interlayer diffusion into the adjacent layer. This mechanism not only provides experimental support for the theoretical prediction of ultrafast relaxation [1,2] and localization [41,42] of the orbital angular momentum but also provides a unified understanding for the absence of orbital current torque in heavy metal/ferromagnet heterostructures [44] and Ti/ferromagnet bilayers [12]. The experimental evidence for strong localization of orbital polarization represents a groundbreaking advance towards solving the orbital torque debate and will immediately impact the blooming field of orbitronics.

**Acknowledgement.** This work is supported partly by the Beijing Natural Science Foundation (Z230006), the National Key Research and Development Program of China (2022YFA1204000), the National Natural Science Foundation of China (12274405), and the Natural Science Foundation of Shaanxi Province (2024JC-YBMS-315).

**Data availability.** The data that support the findings of this study are included in the main text.